\documentclass[final]{siamltex}
\usepackage{amssymb}

\title{Systems of orthogonal polynomials defined by hypergeometric type 
equations\thanks{This work was supported by a grant CERES.}}

\author{Nicolae Cotfas\thanks{Faculty of Physics, University of Bucharest,
PO Box 76-54, Post Office 76, Bucharest, Romania ({\tt ncotfas@yahoo.com}).}}

\begin{document}

\maketitle

\begin{abstract}
A hypergeometric type equation satisfying certain conditions
defines either a finite or an infinite system of orthogonal polynomials. 
We present in a unified and explicit way all these systems of orthogonal 
polynomials, the associated special functions and the corresponding 
raising/lowering operators. This general formalism allows us to extend
some known results to a larger class of functions.
\end{abstract}

\begin{keywords}
orthogonal polynomials, associated special functions, raising operator,
lowering operator, special functions 
\end{keywords}

\begin{AMS}
33C45, 81R05, 81R30
\end{AMS}

\pagestyle{myheadings}
\thispagestyle{plain}
\markboth{N. COTFAS}{SYSTEMS OF ORTHOGONAL POLYNOMIALS}

\section{Introduction}

Many problems in quantum mechanics and
mathematical physics lead to equations of the type
\begin{equation}\label{hypeq}
\sigma (s)y''(s)+\tau (s)y'(s)+\lambda y(s)=0 
\end{equation}
where $\sigma (s)$ and $\tau (s)$ are polynomials of at most second
and first degree, respectively, and $\lambda $ is a constant. 
These equations are usually called {\em equations of hypergeometric
type} \cite{nsu}, and each can be reduced to the self-adjoint form 
\begin{equation}
[\sigma (s)\varrho (s)y'(s)]'+\lambda \varrho (s)y(s)=0 
\end{equation}
by choosing a function $\varrho $ such that 
$[\sigma (s)\varrho (s)]'=\tau (s)\varrho (s)$.

The equation (\ref{hypeq}) is usually considered on an interval $(a,b)$,
chosen such that 
\begin{equation}\begin{array}{r}
\sigma (s)>0\qquad {\rm for\ all}\quad s\in (a,b)\\
\varrho (s)>0\qquad {\rm for\ all}\quad s\in (a,b)\\
\lim_{s\rightarrow a}\sigma (s)\varrho (s)
=\lim_{s\rightarrow b}\sigma (s)\varrho (s)=0.
\end{array}
\end{equation}
Since the form of the equation (\ref{hypeq}) is invariant under a 
change of variable $s\mapsto cs+d$, it is sufficient to analyse the cases
presented in table \ref{table}.
Some restrictions must be imposed on $\alpha $, $\beta $ in
order for the interval $(a,b)$ to exist.

\begin{table}
\caption{The main particular cases} 
\begin{center}
\begin{tabular}{|c|c|l|c|c|}
\hline 
$\begin{array}{l}
\mbox{}\\[-2mm]
\sigma (s)\\[-2mm]
\mbox{}
\end{array}$ & $\tau (s)$  & \mbox{}\qquad \qquad \qquad $\varrho (s)$ & 
$\alpha ,\beta $ &  $(a,b) $\\
\hline 
$\begin{array}{l}
\mbox{}\\[-3mm]
1\\[-3mm]
\mbox{}
\end{array}$ 
& $\alpha s\!+\!\beta $ & ${\rm e}^{\alpha s^2/2+\beta s}$ & $\alpha <0$
& $(-\infty , \infty )$\\ \hline 
$\begin{array}{l}
\mbox{}\\[-3mm]
s\\[-3mm]
\mbox{}
\end{array}$ 
& $\alpha s\!+\!\beta $ & $s^{\beta -1} {\rm e}^{\alpha s}$ & 
$\alpha \!<\!0$, $\beta \!>\!0$& $(0,\infty )$\\ 
\hline
$\begin{array}{l}
\mbox{}\\[-3mm]
1\!-\!s^2\\[-3mm]
\mbox{}
\end{array}$ 
& $\alpha s\!+\!\beta $ & $(1\!+\!s)^{-(\alpha -\beta )/2-1}
(1\!-\!s)^{-(\alpha +\beta )/2-1}$ & 
$\alpha \!<\! \beta <-\alpha $ & $(-1,1)$\\ 
\hline  
$\begin{array}{l}
\mbox{}\\[-3mm]
s^2\!-\!1\\[-3mm]
\mbox{}
\end{array}$ 
& $\alpha s\!+\!\beta $ & $(s\!+\!1)^{(\alpha -\beta )/2-1}
(s\!-\!1)^{(\alpha +\beta )/2-1}$ &
$-\beta \!<\! \alpha <0$ & $(1,\infty )$\\ 
\hline 
$\begin{array}{l}
\mbox{}\\[-3mm]
s^2\\[-3mm]
\mbox{}
\end{array}$ 
& $\alpha s\!+\!\beta $ & $s^{\alpha -2}{\rm e}^{-\beta /s}$ & $\alpha \!<\!0$, 
$\beta \!>\!0$ & $(0,\infty )$\\
\hline 
$\begin{array}{l}
\mbox{}\\[-3mm]
s^2\!+\!1\\[-3mm]
\mbox{}
\end{array}$ 
& $\alpha s\!+\!\beta $ & $(1+s^2)^{\alpha /2-1}{\rm e}^{\beta \arctan s}$ & 
$\alpha <0$ & $(-\infty ,\infty )$\\
\hline
\end{tabular}
\end{center}
\label{table}
\end{table}

The equation (\ref{hypeq}) defines either a finite or an infinite system of 
orthogonal polynomials depending on the set of all $\gamma \in \mathbb{R}$ for which 
\[ \lim_{s\rightarrow a}\sigma (s)\varrho (s)s^\gamma 
=\lim_{s\rightarrow b}\sigma (s)\varrho (s)s^\gamma =0.\]
A unified view on all the systems of orthogonal polynomials defined by (\ref{hypeq})
was presented in \cite{c}. We think that certain results known in particular cases 
can be extended to a larger class of functions by using this general formalism, 
and our aim is to present some attempts in this direction.

The literature discussing special function theory and its application to mathematical
and theoretical physics is vast, and there are a multitude of different conventions
concerning the definition of functions. Since the expression of the raising/lowering
operators depends directly on the normalizing condition we use, a unified approach 
is not possible without a unified definition for the associated special functions.
Our results are based on a definition presented in section 2. The table \ref{table} allows 
one to pass in each case from our parameters $\alpha $, $\beta $ to the parameters
used in different approach. For classical polynomials we use the definitions from
\cite{nsu}.

\section{Orthogonal polynomials and associated special functions}

In this section we review certain results concerning the systems of orthogonal polynomials 
defined by equation (\ref{hypeq}) and the corresponding associated special functions.
It is well-known \cite{nsu} that for $\lambda =\lambda _l$, where
\begin{equation}
\lambda _l=-\frac{\sigma ''(s)}{2}l(l-1)-\tau '(s)l\qquad  l\in \mathbb{N}
\end{equation}
the equation (\ref{hypeq}) admits a polynomial solution 
$\Psi _l=\Psi _l^{(\alpha ,\beta )}$ of at most $l$ degree
\begin{equation} \label{eq3}
\sigma (s) \Psi _l ''+\tau (s) \Psi _l '+\lambda _l\Psi _l=0.
\end{equation}
If the degree of the polynomial $\Psi _l$ is $l$ then it satisfies the
Rodrigues formula
\begin{equation}
\Psi _l(s)=\frac{B_l}{\varrho (s)}\frac{{\rm d}^l}{{\rm d}s^l}[\sigma ^l(s)\varrho (s)]
\end{equation}
where $B_l$ is a constant. We do not impose any normalizing condition.
Each polynomial $\Psi _l$ is defined only up to a multiplicative constant. 
One can remark that
\begin{equation}\label{gamma1}
\lim_{s\rightarrow a}\sigma (s)\varrho (s)s^\gamma 
=\lim_{s\rightarrow b}\sigma (s)\varrho (s)s^\gamma =0\qquad 
{\rm for }\quad \gamma \in [0,\infty ) 
\end{equation}
in the case $\sigma (s)\in \{ 1,\ s,\ 1-s^2\}$, and 
\begin{equation}\label{gamma2}
\lim_{s\rightarrow a}\sigma (s)\varrho (s)s^\gamma 
=\lim_{s\rightarrow b}\sigma (s)\varrho (s)s^\gamma =0\qquad 
{\rm for }\quad \gamma \in [0,-\alpha )
\end{equation}
in the case $\sigma (s)\in \{ s^2-1,\ s^2,\ s^2+1\}$.
Let 
\begin{equation}
\Lambda =\left\{ \begin{array}{lcl}
\infty & {\rm for} & \sigma (s)\in \{ 1,\ s,\ 1-s^2\}\\[2mm]
\frac{1-\alpha }{2} & {\mbox{}\quad \rm for \quad \mbox{}} & 
\sigma (s)\in \{ s^2-1,\ s^2,\ s^2+1\} .
\end{array}\right.
\end{equation}

\begin{proposition}[\cite{nsu,c}]
a) $\{\Psi _l\ |\ l<\Lambda \}$ is a system of polynomials orthogonal 
with weight function $\varrho (s)$ in $(a,b)$. \\[2mm]
b) \ $\Psi _l$ is a
polynomial of degree $l$ for any $l<\Lambda $.\\[2mm]
c) The function $\Psi _l(s)\sqrt{\varrho (s)}$
is square integrable on $(a,b)$ for any $l<\Lambda $.\\[2mm]
d) A three term recurrence relation
\[ s\Psi _l(s)=\alpha _l \Psi _{l+1}(s)+\beta _l\Psi _l(s)
+\gamma _l\Psi _{l-1}(s) \]
is satisfied for $1<l+1<\Lambda $.\\[2mm]
e) The zeros of $\Psi _l$ are simple and lie in the interval $(a,b)$,
for any $l<\Lambda $. 
\end{proposition}

The polynomials $\Psi _l^{(\alpha ,\beta )}$ can be expressed in terms of the 
classical orthogonal polynomials but in certain cases the relation is not very simple.

\begin{proposition}[\cite{c}]
Up to a multiplicative constant
\begin{equation}\label{classical}
  \Psi _l^{(\alpha ,\beta )}(s)=\left\{ \begin{array}{lcl}
H_l\left(\sqrt{\frac{-\alpha }{2}}\, s-\frac{\beta }{\sqrt{-2\alpha }}\right)  
& {\mbox{}\quad {\rm if}\quad \mbox{}} & \sigma (s)=1\\[2mm]
L_l^{\beta -1}(-\alpha s)  & {\rm if} & \sigma (s)=s\\[2mm]
P_l^{(-(\alpha +\beta )/2-1,\ (-\alpha +\beta )/2-1)}(s)  & {\rm if} & \sigma (s)=1-s^2\\[2mm]
P_l^{((\alpha -\beta )/2-1,\ (\alpha +\beta )/2-1)}(-s)  & {\rm if} & \sigma (s)=s^2-1\\[2mm]
\left(\frac{s}{\beta }\right)^lL_l^{1-\alpha -2l}\left(\frac{\beta }{s}\right) 
& {\rm if} & \sigma (s)=s^2\\[2mm]
{\rm i}^lP_l^{((\alpha +{\rm i}\beta )/2-1,\ (\alpha -{\rm i}\beta )/2-1)}({\rm i}s) 
& {\rm if} & \sigma (s)=s^2+1
\end{array} \right.
\end{equation}
where $H_n$, $L_n^p $ and $P_n^{(p,q)}$ are the Hermite,
Laguerre and Jacobi polynomials, respectively.
\end{proposition}

Let $l\in \mathbb{N}$, $l<\Lambda $, and let $m\in \{ 0,1,...,l\}$.
By differentiating the equation (\ref{eq3}) $m$ times we obtain 
the equation satisfied by the polynomials 
$\psi _{l,m}=\frac{{\rm d}^m}{{\rm d}s^m}\Psi _l$, namely
\begin{equation}\label{varphi}
 \sigma (s)\psi ''_{l,m}
+[\tau (s)+m\sigma '(s)]\psi '_{l,m}
+(\lambda _l-\lambda _m) \psi _{l,m}=0.
\end{equation}
This is an equation of hypergeometric type, and we can write 
it in the self-adjoint form 
\begin{equation}
[\sigma (s) \varrho _m(s)\psi '_{l,m}]'
+(\lambda _l-\lambda _m)\varrho _m(s)\psi _{l,m} =0
\end{equation}
by using the function  $\varrho _m(s)=\sigma ^m(s)\varrho (s)$.
\begin{definition}
The functions 
\begin{equation}\label{def}
\Psi _{l,m}(s)=\kappa ^m(s)\frac{{\rm d}^m}{{\rm d}s^m}\Psi _l(s) 
\qquad where\quad \kappa (s)=\sqrt{\sigma (s)}  
\end{equation}
$l\in \mathbb{N}$, $l<\Lambda $ and $m\in \{ 0,1,...,l\}$,
are called the {\em associated special functions}.
\end{definition}

The equation (\ref{varphi}) multiplied by $\kappa ^m(s)$ 
can be written as
\begin{equation}\label{Hm}
{\bf H}_m \Psi _{l,m}=\lambda _l\Psi _{l,m}
\end{equation}
where ${\bf H}_m$ is the differential operator
\[
{\bf H}_m =-\sigma (s) \frac{d^2}{ds^2}-\tau (s) \frac{d}{ds}
+\frac{m(m-2)}{4}\frac{(\sigma '(s))^2}{\sigma (s)} \]   
\begin{equation}
\label{defHm}
 + \frac{m\tau (s)}{2}\frac{\sigma '(s)}{\sigma (s)}
-\frac{1}{2}m(m-2)\sigma ''(s)-m\tau '(s) .
\end{equation}
\begin{proposition}[\cite{c}]
a) For each $m<\Lambda $, the functions
$\Psi _{l,m}$ with $m\leq l<\Lambda $ 
are orthogonal with weight function $\varrho (s)$ in $(a,b)$.\\[2mm] 
b) $\Psi _{l,m}(s)\sqrt{\varrho (s)}$
is square integrable on $(a,b)$ for $0\leq m\leq l<\Lambda $.\\[2mm]
c) The three term recurrence relation
\begin{equation}
   \mbox{}\qquad \qquad \Psi _{l,m+1}(s) + \left( \frac{\tau (s)}{\kappa (s)}
+2(m-1)\kappa '(s)\right)\Psi _{l,m}(s) 
+(\lambda _l-\lambda _{m-1}) \Psi _{l,m-1}(s)=0 \label{rec}
\end{equation}
is satisfied for any $l<\Lambda $ and any $m\in \{ 1,2,...,l-1\}$.
In addition, we have
\begin{equation}\label{rec1}
\left( \frac{\tau (s)}{\kappa (s)}+
2(l-1)\kappa '(s)\right) \Psi _{l,l}(s)
+(\lambda _l-\lambda _{l-1})\Psi _{l,l-1}(s)=0.
\end{equation}
\end{proposition}
 
For any $l\in \mathbb{N}$, $l<\Lambda $ and any $m\in \{ 0,1,...,l-1\}$, 
by differentiating (\ref{def}), we obtain
\[ \frac{{\rm d}}{{\rm d}s}\Psi _{l,m}(s)
=m\kappa ^{m-1}(s)\kappa '(s)\frac{{\rm d}^m}{{\rm d}s^m}\Psi _l
+\kappa ^m(s)\frac{{\rm d}^{m+1}}{{\rm d}s^{m+1}}\Psi _l(s) \]
that is, the relation 
\[ \frac{{\rm d}}{{\rm d}s}\Psi _{l,m}(s)=m\frac{\kappa '(s)}{\kappa (s)}\Psi _{l,m}(s)+
\frac{1}{\kappa (s)}\Psi_{l,m+1}(s) \]
which can be written as
\begin{equation}\label{Am}
\left(\kappa (s)\frac{d}{ds}-
m\kappa '(s)\right) \Psi _{l,m}(s)=\Psi _{l,m+1}(s).
\end{equation}

If $m\in \{ 1,2,...,l-1\}$ then by substituting (\ref{Am}) into
(\ref{rec}) we get 
\[   \left( \kappa (s)\frac{d}{ds}+\frac{\tau (s)}{\kappa (s)}
+(m-2)\kappa '(s)\right)\Psi _{l,m}(s) 
+(\lambda _l-\lambda _{m-1}) \Psi _{l,m-1}(s)=0 \]
that is,
\begin{equation}\label{Am+}
  \left(-\kappa (s)\frac{d}{ds}-
   \frac{\tau (s)}{\kappa (s)}-(m-1)\kappa '(s)\right)
\Psi _{l,m+1}(s)=(\lambda _l-\lambda _m)\Psi _{l,m}(s).
\end{equation}
for all $m\in \{ 0,1,...,l-2\}$. From (\ref{rec1}) it follows that this 
relation is also satisfied for $m=l-1$.

The relations (\ref{Am})  and (\ref{Am+}) suggest we should consider 
the first order differential operators \cite{jf,c0}
\begin{equation}
  A_m=\kappa (s)\frac{d}{ds}-m\kappa '(s)\qquad
A_m^+=-\kappa (s)\frac{d}{ds}-\frac{\tau (s)}{\kappa (s)}-(m-1)\kappa '(s)
\end{equation}
for $m+1<\Lambda $.
\begin{proposition}[\cite{ih,cks,jf,c}]
 We have 
 \begin{equation}\label{AmAm+}
\mbox{}\quad \qquad A_m\Psi _{l,m}=\Psi _{l,m+1}\qquad 
A_m^+\Psi _{l,m+1}=(\lambda _l\!-\!\lambda _m)\Psi _{l,m}\qquad for \ \ 0\leq m<l< \Lambda .
\end{equation}
\begin{equation}\label{philm}
\mbox{}\qquad \Psi _{l,m}=
\frac{A_m^+ }{\lambda _l-\lambda _m}
\frac{A_{m+1}^+ }{\lambda _l-\lambda _{m+1}}...
\frac{A_{l-1}^+ }{\lambda _l-\lambda _{l-1}}\Psi _{l,l}
\qquad for \ \  0\leq m<l< \Lambda .
\end{equation}
\begin{equation} \label{norm}
||\Psi _{l,m+1}||
=\sqrt{\lambda _l-\lambda _m}\, ||\Psi _{l,m}||\qquad  
for\quad  0\leq m<l< \Lambda .
\end{equation}
\begin{equation}\label{fact}
{\bf H}_m-\lambda _m=A_m^+A_m\qquad {\bf H}_{m+1}-\lambda _m
=A_mA_m^+ \qquad for\quad m+1<\Lambda 
\end{equation} 
\begin{equation}\label{interw}
{\bf H}_mA_m^+=A_m^+{\bf H}_{m+1}\qquad A_m{\bf H}_m
={\bf H}_{m+1}A_m\qquad for \quad m+1<\Lambda .
\end{equation}
\end{proposition}

From (\ref{AmAm+}), (\ref{philm}) and (\ref{norm}) it follows that the 
{\em normalized associated special functions}
\begin{equation}
\tilde \Psi _{l,m}=\frac{\Psi _{l,m}}{||\Psi _{l,m}||}
\end{equation}
satisfy the relations
\begin{eqnarray}\label{relations} 
A_m\tilde \Psi _{l,m}&=&\sqrt{\lambda _l-\lambda _m}\tilde \Psi _{l,m+1} \nonumber \\
A_m^+\tilde \Psi _{l,m+1}&=&\sqrt{\lambda _l-\lambda _m}\tilde \Psi _{l,m}\\
\tilde \Psi _{l,m}&=&
\frac{A_m^+ }{\sqrt{\lambda _l-\lambda _m}}
\frac{A_{m+1}^+ }{\sqrt{\lambda _l-\lambda _{m+1}}}...
\frac{A_{l-1}^+ }{\sqrt{\lambda _l-\lambda _{l-1}}}\tilde \Psi _{l,l}. \nonumber
\end{eqnarray}

\section{A group theoretical approach based on projection method}

The system of functions \ $\tilde \Psi _{l,m}$ \ is the projection of the 
system of functions \cite{a}
\begin{equation}\label{deff}
|l,m):(a,b)\times [-\pi ,\pi ]\longrightarrow \mathbb{C}\qquad 
|l,m)={\rm e}^{{\rm i}m\varphi }\tilde \Phi _{l,m}
\end{equation}
orthogonal with respect to the scalar product
\begin{equation}
\langle F,G\rangle 
=\frac{1}{\sqrt{2\pi }}  \int_{-\pi}^\pi \int_a^b\overline{F(s,\varphi )}\, 
G(s,\varphi )\, \varrho (s){\rm d}s{\rm d}\varphi .
\end{equation}
More exactly, we can identify each function $\tilde \Psi _{l,m}$ with the restriction 
of $|l,m)$ to the subset $(a,b)\times \{ 0\}$.
By using the relation  
\begin{equation}
\frac{\partial }{\partial \varphi }|l,m)={\rm i}m\, |l,m).
\end{equation}
obtained directly from definition (\ref{deff}), and (\ref{relations}) we get
\begin{equation}\begin{array}{l}
{\rm e}^{{\rm i}\varphi }\left( \kappa \frac{\partial }{\partial s}+
{\rm i} \kappa '\frac{\partial }{\partial \varphi }\right) |l,m)=
\sqrt{\lambda _l-\lambda _m}\, |l,m+1)\\[3mm]
{\rm e}^{-{\rm i}\varphi }\left( -\kappa \frac{\partial }{\partial s}+
        {\rm i} \kappa '\frac{\partial }{\partial \varphi }
         -\frac{\tau }{\kappa }+2\kappa '\right) |l,m+1)
=\sqrt{\lambda _l-\lambda _m}\, |l,m).\end{array}
\end{equation}
These relations suggest we should consider the first order differential operators 
\begin{equation}\label{diffop}
\begin{array}{l}
L_+={\rm e}^{{\rm i}\varphi }\left( \kappa \frac{\partial }{\partial s}+
        {\rm i} \kappa '\frac{\partial }{\partial \varphi }\right) \\[2mm]
L_-={\rm e}^{-{\rm i}\varphi }\left( -\kappa \frac{\partial }{\partial s}+
        {\rm i} \kappa '\frac{\partial }{\partial \varphi }
         -\frac{\tau }{\kappa }+2\kappa '\right)\\[2mm]
L_0=-{\rm i}\frac{\partial }{\partial \varphi }
\end{array}
\end{equation}
satisfying the relations 
\[ \begin{array}{l}
L_+|l,m)=\sqrt{\lambda _l-\lambda _m}\, |l,m+1)\\[2mm]
L_-|l,m)=\sqrt{\lambda _l-\lambda _{m-1}}\, |l,m-1) \\[2mm]
L_0|l,m)=m|l,m). \end{array}
\]

One can remark that $L_+|l,l\rangle =0$ and 
\begin{equation}
|l,m)=\frac{1}{\sqrt{\lambda _l-\lambda _m}}
\frac{1}{\sqrt{\lambda _l-\lambda _{m+1}}}\cdots
\frac{1}{\sqrt{\lambda _l-\lambda _{l-1}}}(L_-)^{l-m}|l,l)
\end{equation}
for all $m\in \{ 0,1,2,...,l-1\}$, but,  generally, \ $L_-|l,0)\not =0$.
For example, in the case of Legendre polynomials $\kappa (s)=\sqrt{1-s^2}$, \ $\tau (s)=-2s$
and
\begin{equation}
\begin{array}{l}
L_+={\rm e}^{{\rm i}\varphi }\left( \kappa \frac{\partial }{\partial s}+
        {\rm i} \kappa '\frac{\partial }{\partial \varphi }\right) \\[2mm]
L_-={\rm e}^{-{\rm i}\varphi }\left( -\kappa \frac{\partial }{\partial s}+
        {\rm i} \kappa '\frac{\partial }{\partial \varphi }\right)=-
       \overline{{\rm e}^{{\rm i}\varphi }\left( \kappa \frac{\partial }{\partial s}+
        {\rm i} \kappa '\frac{\partial }{\partial \varphi }\right) }=-\overline{L_+}
\end{array}
\end{equation}
whence
\begin{equation}
\mbox{}\qquad (L_-)^m|l,0)=(-1)^m{\rm e}^{-{\rm i}m\varphi }\Psi _{l,m}
=(-1)^m\overline{|l,m)}\qquad {\rm for\ all\ \ } 
m\in \{ 1,2,...,l\} 
\end{equation}
and $(L_-)^{l+1}|l,0)=0$. The $(2l+1)$-dimensional vector space spannied by 
the set $\{ \ (L_-)^q|l,l)\ |\ q\in \{ 0,1,2,...,2l\} \ \}$
is invariant under the action of $L_+$, $L_-$ and $L_0$. 

The operators defined by (\ref{diffop}) satisfy the relations
\begin{equation}
[L_0,L_\pm]=\pm L_\pm 
\end{equation}
and
\begin{equation}
\mbox{}\qquad  \begin{array}{rl}
[L_+,L_-] & =(-\tau '+2\kappa \kappa ''+2{\kappa '}^2)\mathbb{I}
+{\rm i}(2\kappa \kappa ''+2{\kappa '}^2)\frac{\partial }{\partial \varphi }\\[3mm] 
 & =\left\{ \begin{array}{rcl}
 -\alpha \mathbb{I} & {\mbox{}\quad {\rm for}\quad \mbox{}} & \sigma (s)\in \{ 1,\ s\}\\[2mm]
  2\left(L_0-\frac{\alpha +2}{2}\mathbb{I}\right)& {\rm for} & \sigma (s)=1-s^2\\[2mm]
  -2\left(L_0+\frac{\alpha -2}{2}\mathbb{I}\right)& {\rm for} & 
\sigma (s)\in \{ s^2-1,\ s^2,\ s^2+1\} .
   \end{array} \right. 
\end{array} 
\end{equation}
The Lie algebra $\mathcal{L}$ generated by $L_+$ and $L_-$ is finite dimensional.

\begin{theorem}
\[ 
\mathcal{L}\ \ is\ \ isomorphic\ \ to \left\{
\begin{array}{lll}
Heisenberg-Weyl\ \  algebra & if & \sigma (s)\in \{ 1,\, s\} \\[2mm]
su(2) & if & \sigma (s)=1-s^2 \\[2mm]
su(1,1) & if & \sigma (s)\in \{ s^2\!-\!1,\, s^2,\, s^2\!+\!1\}
\end{array} \right. 
\]
\end{theorem}
{\em Proof.} 
If $\sigma (s)\in \{ 1,\, s\}$ then the operators 
$K_+\!=\!\sqrt{-1/\alpha }\, L_+$ and $K_-\!=\!-\sqrt{-1/\alpha }\, L_-$ 
satisfy the relations
\[ [K_+,K_-]=-\mathbb{I}\qquad [\mathbb{I},K_\pm ]=0.\]
In the case $\sigma (s)=1-s^2$ the operators 
$K_+=L_+$, \  $K_-=L_-$ \ and \ $K_0=L_0-\frac{\alpha +2}{2}\mathbb{I}$
satisfy the relations
\[ [K_+,K_-]=2K_0\qquad [K_0,K_\pm ]=\pm K_\pm .\]
If $\sigma (s)\in \{ s^2\!-\!1,\, s^2,\, s^2\!+\!1\}$ the operators
$K_+=L_+$, \ $K_-=L_-$ \ and \ $K_0=L_0+\frac{\alpha -2}{2}\mathbb{I}$
satisfy the relations
\[ [K_+,K_-]=-2K_0\qquad [K_0,K_\pm ]=\pm K_\pm  .\qquad \endproof \]

In the case $\sigma (s)=1-s^2$, the functions $|l,m)$ satisfy the relations
\begin{equation}\mbox{}\qquad \begin{array}{llll}
K_0|l,m)\!=\! (\Phi +m-l)\, |l,m) & {\rm for}& m\!\in \! \{ 0,1,...,l\}\\[2mm]
K_+|l,m)\!=\!  \sqrt{(l-m)(l+m-\alpha -1)}\, |l,m+1) & 
{\rm for} & m\!\in \!\{ 0,1,...,l-1\}\\[2mm]
K_-|l,m)\!=\!  \sqrt{(l-m+1)(l+m-\alpha -2)}\, |l,m-1) & 
{\rm for} & m\!\in \!\{ 1,2,...,l\}\\[2mm]
C|l,m)  \!=\!  \Phi (\Phi +1)\, |l,m) & {\rm for} & m\!\in \!\{ 0,1,...,l\}
\end{array}
\end{equation}
where $C=K_-K_++K_0(K_0+\mathbb{I})$ is the Casimir operator of $su(2)$ and 
$\Phi =l-\frac{\alpha }{2}-1$.

In the case $\sigma (s)\in \{ s^2-1,\, s^2,\, s^2+1\}$, 
the functions $|l,m)$ satisfy the relations
\begin{equation}\mbox{}\qquad \begin{array}{llll}
K_0|l,m)\!=\! (\Phi +m-l)\, |l,m) & {\rm for}& m\!\in \! \{ 0,1,...,l\}\\[2mm]
K_+|l,m)\!=\!  \sqrt{(m-l)(m+l+\alpha -1)}\, |l,m+1) & 
{\rm for} & m\!\in \!\{ 0,1,...,l-1\}\\[2mm]
K_-|l,m)\!=\!  \sqrt{(m-l-1)(m+l+\alpha -2)}\, |l,m-1) & 
{\rm for} & m\!\in \!\{ 1,2,...,l\}\\[2mm]
C|l,m)  \!=\!  -\Phi (\Phi +1)\, |l,m) & {\rm for} & m\!\in \!\{ 0,1,...,l\}
\end{array}
\end{equation}
where $C=K_-K_+-K_0(K_0+\mathbb{I})$ is the Casimir operator of $su(1,1)$ and
$\Phi =l+\frac{\alpha }{2}-1$.

\section{Some systems of coherent states}

\begin{figure}
\setlength{\unitlength}{1mm}
\begin{picture}(100,55)(-5,0)
\put(18.7,7){$a_0 $}
\put(11.8,7){$a_0^+$}
\put(28,18.3){$A_0^+$}
\put(28,12){$A_0$}
\put(25.2,8){$U_0$}
\put(18.7,22){$a_0 $}
\put(11.8,22){$a_0^+$}
\put(28,33.3){$A_0^+$}
\put(28,27){$A_0$}
\put(25.2,23){$U_0$}
\put(18.7,37){$a_0 $}
\put(11.8,37){$a_0^+$}
\put(28,48.3){$A_0^+$}
\put(28,42){$A_0$}
\put(25.2,38){$U_0$}
\put(43.7,22){$a_1 $}
\put(36.8,22){$a_1^+$}
\put(53,33.3){$A_1^+$}
\put(53,27){$A_1$}
\put(50.2,23){$U_1$}
\put(43.7,37){$a_1 $}
\put(36.8,37){$a_1^+$}
\put(53,48.3){$A_1^+$}
\put(53,42){$A_1$}
\put(50.2,38){$U_1$}
\put(68.7,37){$a_2 $}
\put(61.8,37){$a_2^+$}
\put(78,48.3){$A_2^+$}
\put(78,42){$A_2$}
\put(75.2,38){$U_2$}
\put(16,52){$.$}
\put(16,51){$.$}
\put(16,50){$.$}
\put(41,52){$.$}
\put(41,51){$.$}
\put(41,50){$.$}
\put(66,52){$.$}
\put(66,51){$.$}
\put(66,50){$.$}
\put(91,52){$.$}
\put(91,51){$.$}
\put(91,50){$.$}
\put(14.5,0){$\tilde \Psi _{0,0}$}
\put(14.5,15){$\tilde \Psi _{1,0}$}
\put(14.5,30){$\tilde \Psi _{2,0}$}
\put(14.5,45){$\tilde \Psi _{3,0}$}
\put(39.5,15){$\tilde \Psi _{1,1}$}
\put(39.5,30){$\tilde \Psi _{2,1}$}
\put(39.5,45){$\tilde \Psi _{3,1}$}
\put(64.5,30){$\tilde \Psi _{2,2}$}
\put(64.5,45){$\tilde \Psi _{3,2}$}
\put(89.5,45){$\tilde \Psi _{3,3}$}
\put(16,4){\vector(0,1){9}}
\put(18,13){\vector(0,-1){9}}
\put(22,15){\vector(1,0){17}}
\put(39,17){\vector(-1,0){17}}
\put(22,3){\vector(3,2){16}}
\put(16,19){\vector(0,1){9}}
\put(18,28){\vector(0,-1){9}}
\put(22,30){\vector(1,0){17}}
\put(39,32){\vector(-1,0){17}}
\put(22,18){\vector(3,2){16}}
\put(16,34){\vector(0,1){9}}
\put(18,43){\vector(0,-1){9}}
\put(22,45){\vector(1,0){17}}
\put(39,47){\vector(-1,0){17}}
\put(22,33){\vector(3,2){16}}
\put(41,19){\vector(0,1){9}}
\put(43,28){\vector(0,-1){9}}
\put(47,30){\vector(1,0){17}}
\put(64,32){\vector(-1,0){17}}
\put(47,18){\vector(3,2){16}}
\put(41,34){\vector(0,1){9}}
\put(43,43){\vector(0,-1){9}}
\put(47,45){\vector(1,0){17}}
\put(64,47){\vector(-1,0){17}}
\put(47,33){\vector(3,2){16}}
\put(66,34){\vector(0,1){9}}
\put(68,43){\vector(0,-1){9}}
\put(72,45){\vector(1,0){17}}
\put(89,47){\vector(-1,0){17}}
\put(72,33){\vector(3,2){16}}
\end{picture}
\caption{The operators $A_m$, $A_m^+$, $a_m$, $a_m^+$ and $U_m$ relating the 
functions $\tilde \Psi _{l,m}$.}
\label{figure}
\end{figure}
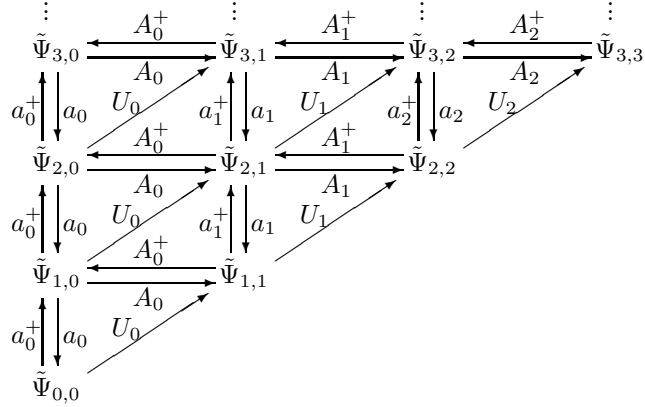

In this section we restrict us to the case $\sigma (s)\!\in \!\{ 1, s, 1\!-\!s^2\}.$ 
For each $m\in \mathbb{N}$, the sequence
$\tilde \Psi _{m,m} ,\, \tilde \Psi _{m+1,m} ,\, \tilde \Psi _{m+2,m},...$
is an orthonormal basis in the Hilbert space
\begin{equation}
{\cal H}=\left\{ \psi :(a,b)\longrightarrow \mathbb{C}\ \left|
\ \int_a^b|\psi (s)|^2\varrho (s)ds<\infty \right. \right\}
\end{equation}
with the scalar product given by 
\begin{equation}
\langle \psi _1, \psi _2\rangle =\int_a^b\overline{\psi _1(s)}\, 
{\psi _2}(s)\, \varrho (s)ds \, .
\end{equation}
The linear operator defined by (see figure \ref{figure}) 
\begin{equation}
U_m:{\cal H}\longrightarrow {\cal H}\qquad 
U_m \tilde \Psi _{l,m} =\tilde \Psi _{l+1,m+1} 
\end{equation}
is a unitary operator, the operators  
\begin{equation}
a_m=U_m^+A_m\qquad a_m^+=A_m^+U_m
\end{equation}
are mutually adjoint, and
\begin{equation} \begin{array}{lll}\label{repsu11}
a_m\tilde \Psi _{l,m}=\sqrt{\lambda _{l}-\lambda _m}\, \tilde \Psi _{l-1,m} &
{\rm for}  &  l\geq m+1\nonumber \\[2mm]
a_m^+\tilde \Psi _{l,m}=\sqrt{\lambda _{l+1}-\lambda _m}\, \tilde \Psi _{l+1,m} &
{\rm for}  & l\geq m\\[2mm]
\tilde \Psi _{l,m}=\frac{(a_m^+)^{l-m}}{\sqrt{(\lambda _l-\lambda _m)
(\lambda _{l-1}-\lambda _m)...(\lambda _{m+1}-\lambda _m)}}\tilde \Psi _{m,m}& {\rm for}
& l>m .
\end{array}
\end{equation}

Since
\begin{equation}
a_m a_m^+\tilde \Psi _{l,m}=(\lambda _{l+1}-\lambda _m)\tilde \Psi _{l,m}\qquad 
a_m^+a_m \tilde \Psi _{l,m}=(\lambda _{l}-\lambda _m)\tilde \Psi _{l,m}
\end{equation}
we get the factorization 
\begin{equation}
{\bf H}_m-\lambda _m=a_m^+a_m \, 
\end{equation}
and the relation
\begin{equation}\label{Lie}
[a_m,a_m^+ ]\tilde \Psi _{l,m}=(\lambda _{l+1}-\lambda _l)\tilde \Psi _{l,m} .
\end{equation}
By using the operator 
\begin{equation}
R_m:{\cal H}_m\longrightarrow {\cal H}_m\qquad 
R_m\tilde \Psi _{l,m}=\frac{-\sigma ''l-\alpha }{2}\tilde \Psi _{l,m}
\end{equation}
the relation (\ref{Lie}) can be written as
\begin{equation}\label{Rm}
[a_m^+,a_m]=-2R_m .
\end{equation}
Since 
\begin{equation}
[R_m,a_m^+]=-\frac{\sigma ''}{2}a_m^+\qquad [R_m,a_m]=\frac{\sigma ''}{2}a_m 
\end{equation}
it follows that the Lie algebra ${\cal L}_m$ generated by 
$\{ a_m^+,a_m \}$ is finite dimensional.

\begin{theorem}
\[
{\cal L}_m\ is \ isomorphic \ to\ \ 
\left\{ \begin{array}{lcl}
Heisenberg-Weyl\  algebra & for & \sigma (s)\in \{ 1,\ s\}\\[2mm]
su(1,1) & for & \sigma (s)=1-s^2
\end{array} \right.
\]
\end{theorem}

{\em Proof}.
In the case $\sigma (s)\in \{ 1,s\}$ the operator $R_m$ is a constant
operator, namely, $R_m=-\alpha $. Since $\alpha <0$, the operators 
$P_+=\sqrt{-1/\alpha }\, a_m^+$, $P_-=\sqrt{-1/\alpha }\, a_m$ and the 
identity operator $\mathbb{I}$ form a basis of ${\cal L}_m$ such that
\[ [P_+,P_-]=-\mathbb{I}\qquad [\mathbb{I},P_\pm ]=0 \]
that is, ${\cal L}_m$ is isomorphic to the Heisenberg-Weyl algebra 
$h(2)$.\\ If $\sigma (s)=1-s^2$ then  
$K_+=a_m^+$,  $K_-=a_m $ and $K_0=R_m$ form a basis of ${\cal L}_m$
such that
\[ [K_+,K_-]=-2 K_0 \qquad [K_0,K_\pm ]=\pm K_\pm  .\qquad  \endproof \]

In the case $\sigma (s)=1-s^2$, the functions 
$\tilde \Psi _{m,m} ,\, \tilde \Psi _{m+1,m} ,\, \tilde \Psi _{m+2,m},...$, 
satisfy the relations
\begin{eqnarray}
K_0\tilde \Psi _{l,m} & = & \left(l-\frac{\alpha }{2}\right)\, \tilde \Psi _{l,m} \\[2mm]
K_+\tilde \Psi _{l,m} & = & \sqrt{(l-m+1)(l+m-\alpha )}\, \tilde \Psi _{l+1,m} \\[2mm]
K_-\tilde \Psi _{l,m} & = & \sqrt{(l-m)(l+m-1-\alpha )}\, \tilde \Psi _{l-1,m} \\[2mm]
C\tilde \Psi _{l,m} & = & 
-\left( \frac{\alpha }{2}-m\right)\left( \frac{\alpha }{2}-m+1\right)\, \tilde \Psi _{l,m}
\end{eqnarray}
where $C=K_-K_+-K_0(K_0+\mathbb{I})$ is the Casimir operator of $su(1,1)$.
If we denote 
\begin{equation}
E_0=m-\frac{\alpha }{2}=-\Phi \qquad |\Phi , n\rangle =\tilde \Psi _{m+n,m}
\end{equation}
then the above relations can be written as
\begin{eqnarray}
K_0|\Phi ,n\rangle & = & (E_0+n)\, |\Phi ,n\rangle \\[2mm]
K_+|\Phi ,n\rangle  & = & \sqrt{(\Phi +E_0+n+1)(E_0-\Phi +n)}\, |\Phi ,n+1\rangle  \\[2mm]
K_-|\Phi ,n\rangle  & = & \sqrt{(\Phi +E_0+n)(E_0-\Phi +n-1)}\, |\Phi ,n-1\rangle  \\[2mm]
C|\Phi ,n\rangle & = & -\Phi (\Phi +1)\, |\Phi ,n\rangle .
\end{eqnarray}
and show that \cite{bg,w}, in case $\sigma (s)=1-s^2$, the representation of $su(1,1)$ defined by 
(\ref{repsu11}) in $\mathcal{H}$
is the irreducible discrete representation $D^+\left( \frac{\alpha }{2}-m\right)$.

Let $m\in \mathbb{N}$ be a fixed natural number. 
The functions \ $|0\rangle $, $|1\rangle $, $|2\rangle $, $\cdots $ , where 
\begin{equation}
|n\rangle =\tilde \Psi _{m+n,m}
\end{equation}
satisfy the relations
\begin{eqnarray}
a_m|n\rangle & = & \sqrt{e_n}\, |n-1\rangle \nonumber \\[2mm]
a_m^+|n\rangle & = & \sqrt{e_{n+1}}\, |n+1\rangle \\[2mm]
({\bf H}_m-\lambda _m)|n\rangle & = & e_n|n\rangle \nonumber 
\end{eqnarray}
where
\begin{equation}
e_n=\lambda _{m+n}-\lambda _m=\left\{ 
\begin{array}{lcl}
-\alpha n & {\rm if} & \sigma(s)\in \{ 1,s\}\\[2mm]
n(n+2m-\alpha -1) & {\rm if } & \sigma (s)=1-s^2 .
\end{array} \right. 
\end{equation}
Some useful systems of coherent states can be defined \cite{bg} by using these 
relations, the confluent hypergeometric function
\begin{equation}
\mbox{}_0F_1(c;z)=1+\frac{1}{c}\frac{z}{1!}+\frac{1}{c(c+1)}\frac{z^2}{2!}
+\frac{1}{c(c+1)(c+2)}\frac{z^3}{3!}+\cdots 
\end{equation}
and the modified Bessel function
\begin{equation} \mbox{}\qquad 
K_\nu (z)=\frac{\pi }{2}\frac{I_{-\nu }(z)-I_\nu (z)}{{\rm sin}\, (\nu \pi )}
\qquad {\rm where}\qquad 
I_\nu (z)=\sum_{n=0}^\infty \frac{\left(\frac{1}{2}z\right)^{\nu +2n}}
{n!\, \Gamma (\nu +n+1)}.
\end{equation}

\begin{theorem}
a) If $\sigma (s)\in \{ 1,s\}$ then $\{ \ |z\rangle \ |\ z\in \mathbb{C}\}$, where 
\begin{equation}\label{cs1}
|z\rangle ={\rm e}^{\frac{|z|^2}{2\alpha }}\sum_{n=0}^\infty
\frac{z^n}{\sqrt{n!\, (-\alpha )^n}}|n\rangle 
\end{equation}
is a system of coherent states in $\mathcal{H}$ such that
\begin{equation} \mbox{}\qquad 
\langle z|z\rangle =1\qquad a_m|z\rangle =z|z\rangle \qquad and\qquad 
\frac{1}{\pi \sqrt{-\alpha }}\int_\mathbb{C} d({\rm Re}z)\, 
d({\rm Im}z)|z\rangle \langle z|=\mathbb{I}.
\end{equation}
b) If $\sigma (s)=1-s^2$ then $\{ \ |z\rangle \ |\ z\in \mathbb{C}\}$, where 
\begin{equation}\label{cs2}
|z\rangle =\sqrt{\Gamma (2m-\alpha )}
\sum_{n=0}^\infty \frac{z^n}{\sqrt{n!\, \Gamma (n+2m-\alpha )}}|n\rangle 
\end{equation}
is a system of coherent states in $\mathcal{H}$ such that 
\begin{equation}\mbox{}\qquad 
\langle z|z\rangle =\mbox{}_0F_1(2m-\alpha ;|z|^2)\qquad a_m|z\rangle =z|z\rangle 
\quad and \quad \int_\mathbb{C} d\mu (z)\, |z\rangle \langle z|=\mathbb{I}
\end{equation}
where
\begin{equation}
d\mu (z)=\frac{4r^{2m-\alpha }}{\pi \Gamma (2m-\alpha )}
K_{\frac{\alpha +1}{2}-m}(2r)\, dr\, d\theta \qquad and \qquad z=r{\rm e}^{{\rm i}\theta }.
\end{equation}
\end{theorem}
The proof can be found in \cite{bg,ag} (different notations
are used for some parameters).

\end{document}